# Switchable chiral pair density wave in pure $CsV_3Sb_5$

**Authors:** Wei Song[1*], Xiao-Yu Yan[1*], Xin Yu[2*], Desheng Wu[3*], Deng Hu[4,5], Hailang Qin[3], Guowei Liu[3], Hanbin Deng[1], Chao Yan[1]. Muwei Gao[1], Zhiwei Wang[4,5], Rui Wu[2]†, Jia-Xin Yin[1,3]†

**Affiliations:**
[1]Department of Physics and Guangdong Basic Research Center of Excellence for Quantum Science, Southern University of Science and Technology, Shenzhen 518055, China.
[2]Beijing National Laboratory for Condensed Matter Physics, Institute of Physics, Chinese Academy of Sciences, Beijing 100190, China.
[3]Quantum Science Center of Guangdong-Hong Kong-Macao Greater Bay Area, Shenzhen, China.
[4]Centre for Quantum Physics, Key Laboratory of Advanced Optoelectronic Quantum Architecture and Measurement (MOE), School of Physics, Beijing Institute of Technology, Beijing, China.
[5]Beijing Key Lab of Nanophotonics and Ultrafine Optoelectronic Systems, Beijing Institute of Technology, Beijing, China.

**We investigate electron pairing in a super clean kagome superconductor $CsV_3Sb_5$ with a residual resistivity ratio (RRR) of 290. Using the dilution-refrigerator-based scanning tunneling microscopy (STM) at the Synergetic Extreme Condition User Facility (SECUF), we find that the pairing gap exhibits chiral 2×2 modulations, and their chirality can be controlled by magnetic field training. We introduce nonmagnetic impurities to observe the complete suppression of 2×2 pairing modulations in presence of persistent 2×2 charge order. This nonmagnetic pair-breaking effect provides phase-sensitive evidence for pair-density-wave (PDW) induced pairing modulations. Our results support switchable chiral 2×2 PDW in this super clean kagome superconductor.**

The PDW is an unusual pairing state involving finite-momentum Cooper pairs and exhibits spatial modulations of the superconducting order parameter[1]. In the kagome superconductor, the induced chiral 2×2 PDW exhibits several unique features that broadly correspond with a variety of ultra-low temperature and high-resolution experiments as well as advanced theories [2-20]. In the STM study [2,3], the chiral PDW exhibits orbital selectivity, arising from the coupling between uniform superconductivity in the Sb *p*-orbital and the chiral 2×2 charge order in the V *d*-orbital. The tunneling spectrum is gapless [2-5], aligning with thermal transport [6,7], muon spin rotation (μSR) [8] and nuclear quadrupole resonance (NQR) [9]. The absence of uniform pairing in the *d*-orbital is attributed to the time-reversal symmetry-breaking of the chiral charge order within this orbital [10-12]. The gapless *d*-orbital states form residual Fermi arcs in connection with the 2×2 pair modulations, establishing momentum-spatial correspondence for PDW [2,3]. The magnetic field tunability of the PDW chirality [2] agrees with device-based chiral superconducting transport [14,15], both supporting time-reversal symmetry-breaking.

By considering the attractive on-bond pairing interactions and van Hove filling in the kagome lattice, the chiral 2×2 PDW with nontrivial topology can be obtained self-consistently [16]. This theory explains the residual Fermi states detected by transport [5,7], μSR [8], NQR [9] and STM [2-5] as Bogoliubov Fermi arcs, and further predicts the strong pairing gap anisotropy that has recently been confirmed by angle-resolved photoemission spectroscopy (ARPES) data on the Fermi surface with V *d*-orbital character [13]. The Fermi surface with Sb *p*-orbital exhibits isotropic pairing [13], aligning with the STM observations on the orbital selectivity [2,3]. Theoretical analysis suggests that the melting of the chiral PDW could produce charge 6e excitations [16-19] observed in the experiment [20]. Owing to its topological character, the chiral PDW theory [16] predicts quantized anomalous thermal Hall effect in the superconducting state, while a much larger anomalous thermal Hall effect boosted by impurity scattering has recently been detected [21]. Self-consistent calculations [16] have also found that, likely due to the sublattice interference, the gap size modulation measured by the coherence peak-to-peak distance is only a few percent even for a pure PDW state in the V *d*-orbital, which challenges its experimental detection.

While there are reports of switchable chiral transport in $CsV_3Sb_5$ [14,15,22] and STM evidence for field switchable 2×2 PDW in $KV_3Sb_5$ [2], spectroscopic evidence for switching of the chiral PDW in $CsV_3Sb_5$ is still lacking. Broadly speaking, pairing gap modulations measured by STM can be produced by intrinsic PDW or extrinsic mechanisms [23]. The PDW uniquely features phase sign modulations, which makes it sensitive to nonmagnetic impurity scattering [24-26], whereas impurity scattering can promote pairing modulations under certain extrinsic mechanisms [23]. Thus, it is highly desirable to use nonmagnetic impurity scattering as a phase-sensitive probe [27] for PDW in $CsV_3Sb_5$. Recently, we have successfully grown high-quality single crystals of $CsV_3Sb_5$ with an RRR of 290 [Fig. 1(a)], compared to typical values below 100 [28]. The superconducting transition temperature $T_c$, has also increased from 2.5 K [28] to 3.0 K. We studied this crystal by using the dilution-refrigerator-based STM at SECUF. The crystal was cleaved at ~ 10 K and measured at a base temperature of 20 mK. Although the electronic temperature is not accurately determined but the data are comparable to previous measurements at an electronic temperature estimated to be lower than 90 mK [2,3,29].

We focus on the Sb surface that is tightly bonded with the V kagome lattice. On this surface, we detect a charge order gap of $\Delta_{CO}$ = 40 meV [Fig. 1(b)] and a superconducting gap of $\Delta_{SC}$ = 0.45 meV [Fig. 1(c)]. The pairing gap is U shaped, but still with detectable flat in-gap states. The scaling number for $\Delta_{SC}/T_c$ for this high-quality kagome superconductor aligns with previous data in $AV_3Sb_5$ system [2,3] [Fig. 1(d)]. The topographic image and its Fourier transform analysis in Fig. 2(e) reveal the 2×2 charge order and 1×4 stripes. These stripes have not been detected by bulk scattering techniques, and thus are likely of surface origin. The existence of 1×4 stripes makes it harder to resolve the 2×2 vector peaks compared with the data in $KV_3Sb_5$. Therefore, imaging for a large field of view is required.

In the same field of view, we measure the absolute position for coherence peaks at positive and negative energies to determine the gap map $\Delta_{SC}^{\pm}(\boldsymbol{r})$ in Fig. 1(f) and (g), respectively. These two maps closely resemble each other, demonstrating the particle-hole symmetry of the pairing gap. The zoomed in spectrums in Figs. 1(f) and (g) illustrate the tiny energy difference in the gap variations, which is detectable by our high-resolution measurement. The Fourier transforms of the gap maps reveal the pairing modulations of $\Delta_{SC}^{\pm}(\boldsymbol{q})$. We focus on the 2×2 bulk ordering vectors and find that the 2×2 pairing modulations have different intensities for three different directions. If we count from the lower intensity vector peaks to higher intensity vector peaks, we obtain an anticlockwise chirality of the 2×2 pairing modulations for both of the $\Delta_{SC}^{\pm}(\boldsymbol{q})$ maps. We further present pairing map $\Delta_{SC}(r) = [\Delta_{SC}^{+}(r) + \Delta_{SC}^{-}(\boldsymbol{r})]/2$ in Fig. 1(h) and the Fourier transform identifies similar anticlockwise chiral 2×2 pairing modulations. It is a challenging task to observe the 2×2 pairing modulations in the gap map, which is absent in some of previous STM measurements [3,4] mainly because of the relatively smaller mapping area. The detection of the tiny modulations of the pairing gap magnitude in $AV_3Sb_5$ system also requires extremely low electronic temperature, high spatial resolution and high signal to noise ratio [See SI for details].

We further perform the magnetic field training experiment in $CsV_3Sb_5$. We apply the magnetic field perpendicular to the kagome lattice with $B$ = -2T and then withdraw the field back to 0T. Measuring the pairing gap map [Fig. 2(a)-(c)], we find that the chirality of 2×2 pairing modulations is clockwise. Then, we apply the magnetic field to +2T and withdraw the field to 0T. Remeasuring the pairing gap map for the same field of view [Fig. 2(d)-(f)], we find that the chirality of 2×2 pairing modulations is switched to anticlockwise. The pairing gaps are measured at 0T because a tiny magnetic field can destroy superconductivity and the gap feature. The magnetic field switching of the 2×2 chiral pairing modulations aligns with the superconducting chiral transport reported [14,15] for $CsV_3Sb_5$, and supports the time-reversal symmetry breaking that persists in the pairing state.

We also note that there are extra band structure related quasi-particle interference feature in the Fourier transform data of the gap map, this is likely to be related with the impurity induced pair-breaking interference effect as discussed earlier in the time-reversal symmetry-breaking superconducting state [2,23]. A careful examination of all our $\Delta(\boldsymbol{q})$ data suggests that the 4a/3×4a/3 appears to be connected with such quasi-particle interference vectors. In $KV_3Sb_5$, we can detect 4a/3×4a/3 gap modulations only at the defect rich region and cannot detect it in the defect free region both in the $\Delta(\boldsymbol{q})$ map or the Josephson tunneling map. Thus, we attribute it to the quasi-particle interference assisted gap modulations in presence of impurity scattering and time-reversal symmetry breaking of the pairing state.

The order parameter for charge order generally does not feature sign reversal in real space. By contrast, its induced pairing modulation in the superconducting state may feature sign reversal, thus forming a PDW [Fig. 3(a)] if there is no uniform pairing in the same orbital. In kagome superconductors, the broken time-reversal symmetry of the chiral charge order state may substantially reduce the uniform pairing in the V *d*-orbital, and the *d*-orbital may host a true PDW state in the kagome lattice [2,3] that may differ from the induced PDW in other platforms [30-32]. Theoretical analysis has proposed that the PDW will be highly sensitive to nonmagnetic impurity scattering [24-26], and this proposal has recently been realized in nonmagnetic Ta impurity doped $KV_3Sb_5$ by observing the global and local PDW breaking effect [27]. We further provide this phase sensitive evidence for the PDW scenario in $CsV_3Sb_5$ with 7% nonmagnetic isovalent Nb dopants.

The 7% Nb doped $CsV_3Sb_5$ has a reduced charge order and enhanced superconductivity in the transport data [Fig. 3(b)]. The charge order $\rho_Q$ is reduced by 34% and superconducting order is enhanced by 47%. Thus, Ginzburg-Landau theory predicts an enhancement for the induced pairing modulation, as the pairing modulation is related to $\Delta_Q \propto \rho_Q \Delta_{SC}$.

In the STM experiments, the individual Nb dopants can be imaged with a large bias voltage topography in Fig. 3(c), and the counting of these dopants is consistent with 7% nominal doping. Spectroscopically, we also detected reduced charge order gap [Fig. 3(d)], and the 2×2 charge order can be detected by the low-bias topographic data [Fig. 3(e)] and its Fourier transform [Fig. 3(f)]. We detect an enhanced superconducting gap [Fig. 3(g)] that is almost fully gapped. We measure the pairing gap map in Fig. 3(h), and its Fourier transform analysis suggests no detectable signal at the 2×2 vectors in Fig. 3(i), although the modulations at Bragg vectors are still very strong, which confirms that the high spatial and energy resolution of our measurement. The observation of diminished 2×2 pairing modulation is in sharp contrast with the Ginzburg-Landau estimation and the detectable 2×2 charge order, but is consistent with the nonmagnetic PDW-breaking effect.

The observed nonmagnetic PDW-breaking effect agrees well with recent observations in $KV_3Sb_5$ system and explains a paradox in ARPES results. In $KV_3Sb_5$, 4% nonmagnetic Ta impurities reduce the global PDW order by 67% [27]. Assuming the Nb and Ta have a similar pair-breaking effect, the PDW order should readily be destroyed by 6% such impurities, which aligns with our observation at 7%. Early ARPES data [33] also detected isotropic pairing in 7% Nb-doped $CsV_3Sb_5$ and concluded that the existence of charge order would not introduce anisotropic pairing. However, recent ARPES data on pristine $CsV_3Sb_5$ reveals strong anisotropy of the pairing gap on the *d*-orbital pockets. This apparent paradox can now be reconciled by the fact the PDW in the *d*-orbital leads to the gap anisotropy in $CsV_3Sb_5$ [13], while the nonmagnetic PDW breaking effect destroys PDW in 7% Nb doped $CsV_3Sb_5$ resulting in an isotropic pairing gap.

Taken together with previous STM data in $KV_3Sb_5$ and $RbV_3Sb_5$, our new results establish the ubiquitous switchable chiral PDW in $AV_3Sb_5$ systems, which should have broad connections to various ultra-low temperature high-resolution experiments and advanced theories [34-41] on the ground state of kagome superconductors. The methodology applied here can be extended to examine PDW scenarios in other superconducting platforms.

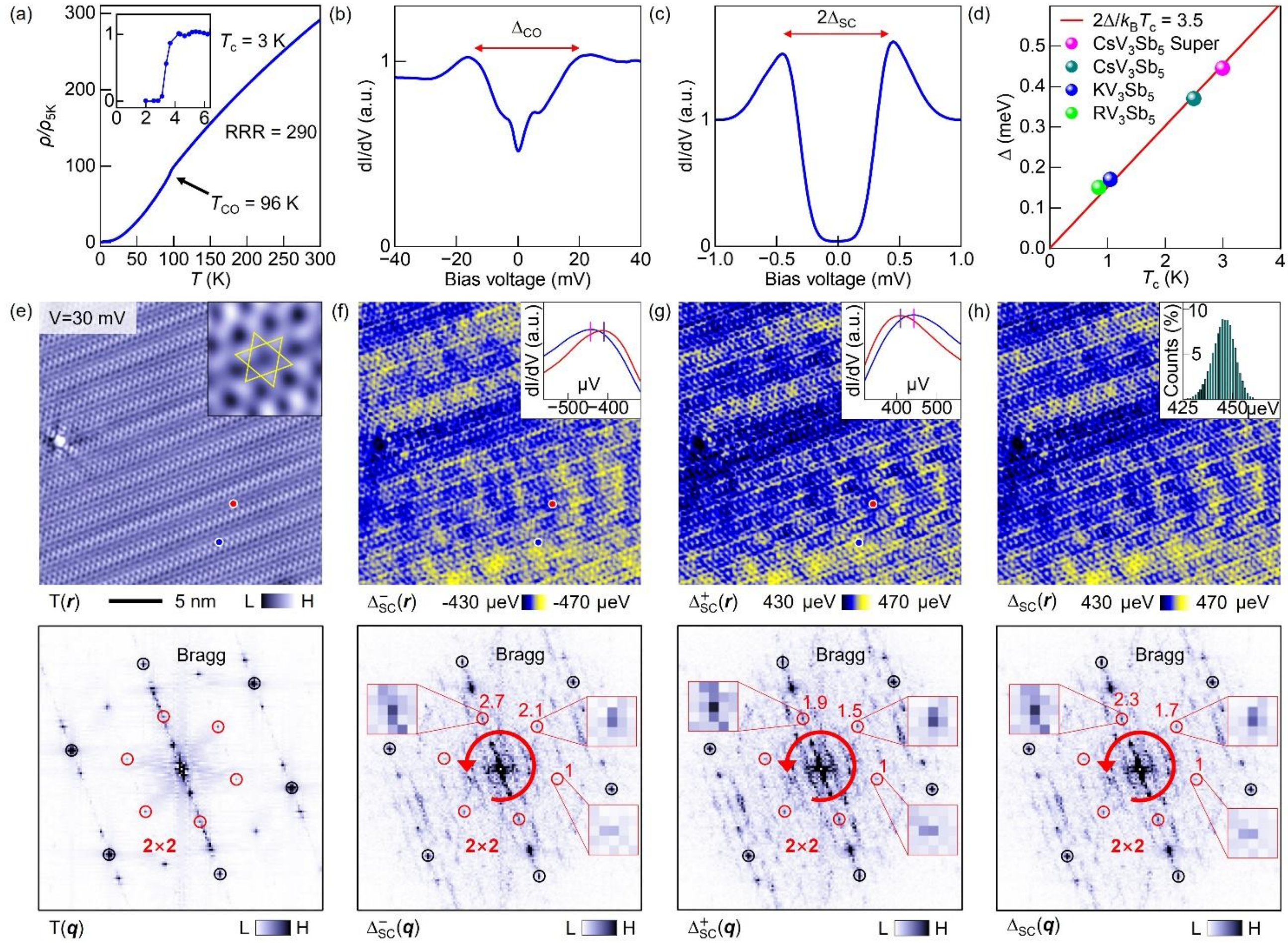

Fig. 1 (a) Resistivity data of the super clean $CsV_3Sb_5$ showing high RRR of 290, a charge order transition at $T_{co}$ = 95 K, and superconducting transition at $T_c$ = 3.0 K (inset). (b) Intermediate-energy tunneling spectrum revealing the charge order gap $\Delta_{CO}$. We use $V$ = 40 mV, $I$ = 1 nA and modulation voltage $V_m$ =1 mV. (c) Low-energy tunneling spectrum revealing the superconducting gap $\Delta_{SC}$. We use $V$ = 2 mV, $I$ = 0.5 nA, $V_m$ =0.05 mV. (d) Scaling analysis between Tc and $\Delta_{SC}$ for various of $AV_3Sb_5$ superconductors. (e) Topographic image of a large Sb surface (upper panel) and its Fourier transform analysis (lower panel). The black circles mark the Bragg peaks, and the red circles mark the 2×2 vector peaks for charge order. The inset in the upper panel shows a zoomed-in image with marked underlying kagome structures. We use $V$ = 30 mV and $I$ = 500 pA. (f) (g) Pairing gap map $\Delta^{\pm}_{SC}(\boldsymbol{r})$ for the same region (upper panel) and its Fourier transform analysis (lower panel). The inset in the upper panel shows two typical spectrums for identifying the gap difference, and their positions are marked on the main panel by colored dots. We zoom in the 2×2 vector peaks in the lower panel to show the anticlockwise chirality of the 2×2 pairing modulations. Their intensities are also noted by numbers. We use $V$ = 2 mV, $I$ = 1 nA, and $V_m$ = 0.01 mV. (h) Pairing gap map $\Delta_{SC}(\boldsymbol{r})$ for the same region (upper panel) and its Fourier transform analysis (lower panel). The inset in the upper panel shows the histogram of the gap distribution. We use $V$ = 2 mV, $I$ = 1 nA, and $V_m$ = 0.01 mV.

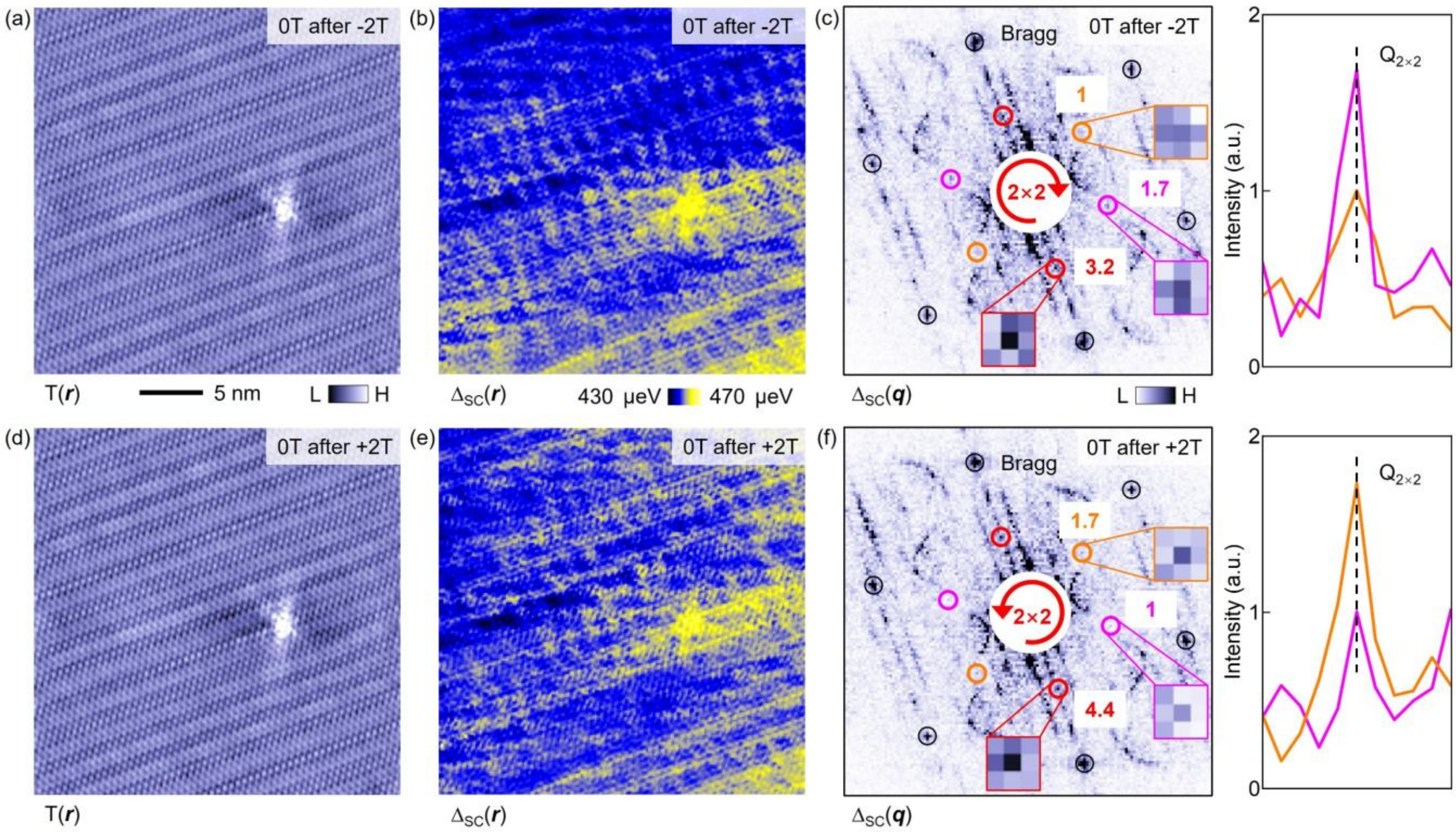

Fig. 2 (a) Topographic image taken at 0T after applying a perpendicular magnetic field of -2T. We use $V = 1$ mV, $I = 1$ nA. (b) Pairing gap map at the same region with -2T training. We use $V = 1$ mV, $I = 1$ nA, and $V_m = 0.01$ mV. (c) Fourier transform of the pairing map showing 2×2 modulation with clockwise chirality, and the right panel plot the line profile across two of the 2×2 vector peaks. (d) Topographic image taken at 0T after applying a perpendicular magnetic field of +2T. We use $V = 1$ mV, $I = 1$ nA, and $V_m = 0.01$ mV. (e) Pairing gap map at the same region with +2T training. We use $V = 1$ mV, $I = 1$ nA, and $V_m = 0.01$ mV. (f) Fourier transform of the pairing map showing 2×2 modulation with anticlockwise chirality, and the right panel plot the line profile across two of the 2×2 vector peaks.

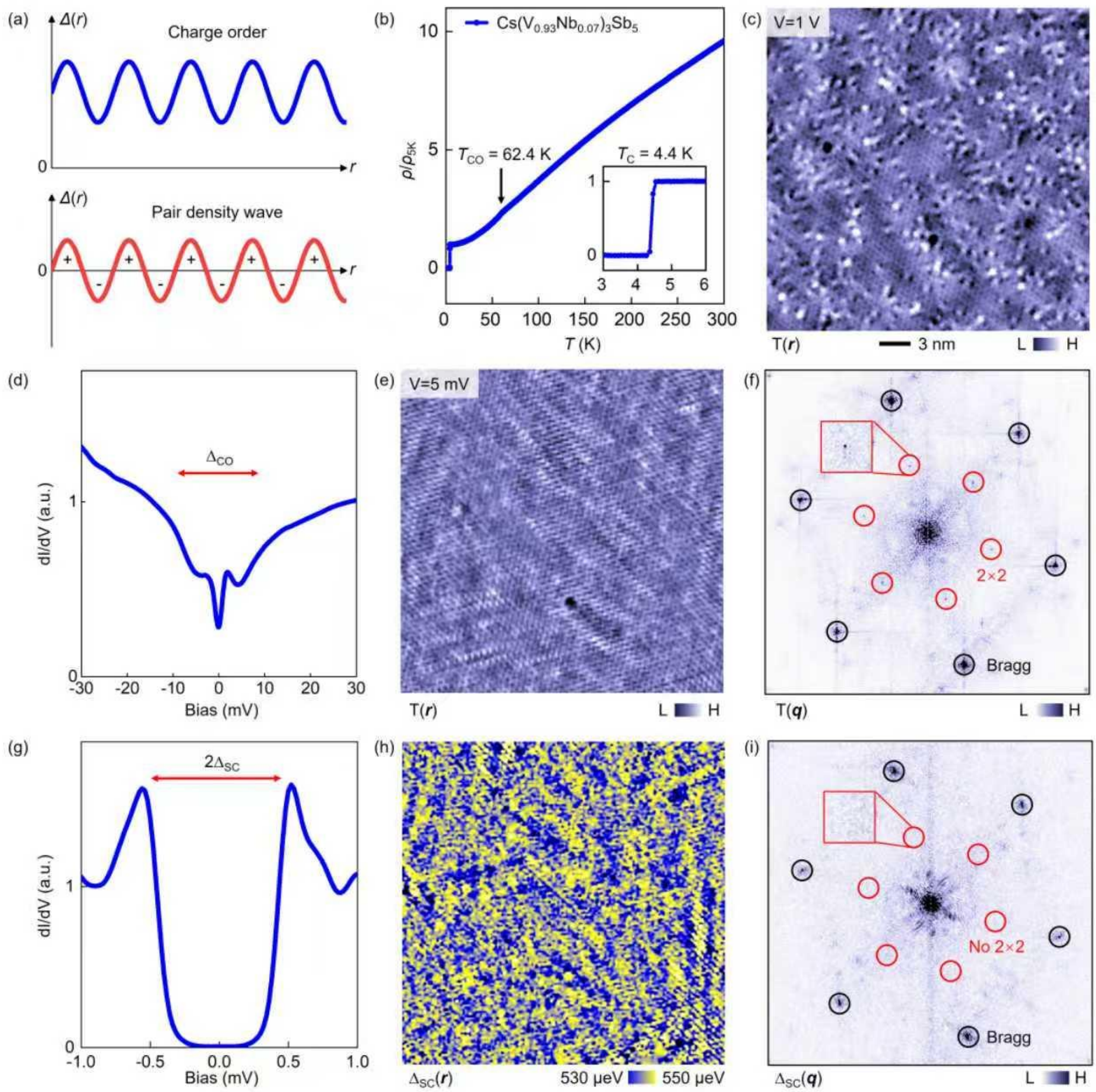

Fig. 3 (a) Schematic showing the order parameter modulations at real space for charge order (upper panel) and PDW (lower panel). The PDW uniquely features sign-reversals that makes it more sensitive to nonmagnetic impurity scattering. (b) Resistivity data of 7% Nb doped $CsV_3Sb_5$ showing a charge order transition at 62.4 K, and superconducting transition at 4.4 K (inset). (c) High-bias topographic image of the Sb surface showing the underlying individual Nb impurities in the kagome lattice. We use $V = 1$ V, $I = 0.5$ nA. (d) Intermediate-energy tunneling spectrum revealing the charge order gap $\Delta_{CO}$. (e) Low-bias topographic image at the same atomic location. We use $V = 5$ mV, $I = 1$ nA, $V_m$ =0.5 mV. (f) Fourier transform of the topographic image showing 2×2 vector peaks. (g) Low-energy tunneling spectrum revealing the superconducting gap $\Delta_{SC}$. We use $V = 2$ mV, $I = 1$ nA, $V_m$ =0.01 mV. (h) Pairing gap map $\Delta_{SC}(\boldsymbol{r})$ for the same region. We use $V = 2$ mV, $I = 1$ nA, $V_m$ =0.01 mV. (i) Fourier transform of the pairing gap showing the absence of 2×2 modulations while the Bragg modulations are still present.

**Acknowledgment**

This work was supported by the Synergetic Extreme Condition User Facility (SECUF, https://cstr.cn/31123.02.SECUF). We acknowledge the support from the National Key R&D Program of China (grant number 2023YFF0718403, 2023YFA1407300), the National Science Foundation of China (grant number 12374060), Guangdong Provincial Quantum Science Strategic Initiative (GDZX2401001).

**Data availability**

Data are available from the authors upon reasonable request.